\documentclass[lettersize,journal]{IEEEtran}
\usepackage{amsmath,amsfonts}
\usepackage{algorithm,algpseudocode}
\usepackage{array}
\usepackage[caption=false,font=footnotesize,labelfont=sf,textfont=sf]{subfig}
\usepackage{textcomp}
\usepackage{stfloats}
\usepackage{url}
\usepackage{verbatim}
\usepackage{graphicx}
\usepackage{cite}
\usepackage{color}
\usepackage{hyperref}
\usepackage{siunitx}

\begin{document}
\title{Hybrid Parallel Collaborative Simulation Framework\\Integrating Device Physics with Circuit Dynamics\\for PDAE-Modeled Power Electronic Equipment }

\author{
    Qingyuan Shi,
    Chijie Zhuang,~\IEEEmembership{Member,~IEEE},   
    Jiapeng Liu,
    Bo Lin, Xiyu Peng, \\Dan Wu, 
    Zhicheng Liu,    
    Rong Zeng,~\IEEEmembership{Senior Member,~IEEE}
   
\thanks{Manuscript received August 15, 2024; revised December 6, 2024. This work is supported by National Natural Science Foundation of China under grant U2166602 and 524B2107. (\textit{Corresponding author: Chijie Zhuang.})

Q. Shi, C. Zhuang, J. Liu, X. Peng, Z. Liu and R. Zeng are with Department of Electrical Engineering, Tsinghua University, Beijing 100084, China (e-mail: shiqy21@mails.tsinghua.edu.cn; chijie@tsinghua.edu.cn; liujiapeng@tsinghua.edu.cn; pengxy22@mails.tsinghua.edu.cn; zc-liu20@mails.tsinghua.edu.cn; zengrong@tsinghua.edu.cn). 

B. Lin, D. Wu are with Beijing Huairou Laboratory, Beijing 101499, China (e-mail: linbo@neps.hrl.ac.cn; wudan@neps.hrl.ac.cn).

Q. Shi and C. Zhuang are also with Beijing Huairou Laboratory, Beijing 101499, China.
}
}

\markboth{}
{Shell \MakeLowercase{\textit{et al.}}: A Sample Article Using IEEEtran.cls for IEEE Journals}

\maketitle

\begin{abstract}
Optimizing high-performance power electronic equipment, such as power converters, requires multiscale simulations that incorporate the physics of power semiconductor devices and the dynamics of other circuit components, especially in conducting Design of Experiments (DoEs),  defining the safe operating area of devices, and analyzing failures related to semiconductor devices. However, current methodologies either overlook the intricacies of device physics or do not achieve satisfactory computational speeds. To bridge this gap, this paper proposes a Hybrid-Parallel Collaborative (HPC) framework specifically designed to analyze the Partial Differential Algebraic Equation (PDAE) modeled power electronic equipment, integrating the device physics and circuit dynamics. The HPC framework employs a dynamic iteration to tackle the challenges inherent in solving the coupled nonlinear PDAE system, and utilizes a hybrid-parallel computing strategy to reduce computing time. Physics-based system partitioning along with hybrid-process-thread parallelization on shared and distributed memory are employed, facilitating the simulation of hundreds of partial differential equations (PDEs)-modeled devices simultaneously without compromising speed. Experiments based on the hybrid line commutated converter and reverse-blocking integrated gate-commutated thyristors are conducted under 3 typical real-world scenarios: semiconductor device optimization for the converter; converter design optimization; and device failure analysis. The HPC framework delivers simulation speed up to 60 times faster than the leading commercial software, while maintaining carrier-level accuracy in the experiments. This speedup becomes more pronounced as the number of semiconductor devices increases. This shows great potential for comprehensive analysis and collaborative optimization of devices and electronic power equipment, particularly in extreme conditions and failure scenarios.
\end{abstract}

\begin{IEEEkeywords}
Power electronics analysis, collaborative simulation, hybrid parallel, PDAE modeling, semiconductor device, device physics, circuit dynamics.
\end{IEEEkeywords}

\vspace{-10pt}

\section{Introduction}
\IEEEPARstart{A}{s} the New Power System, central to China's carbon neutrality solution, continuously evolves, DC technology plays an increasingly critical role in the extensive development of clean power infrastructure and long-distance power transmission. Consequently, an increasing number of high-voltage and high-capacity power electronic equipment, such as power converters, DC circuit breakers, and DC transformers, have been serving as indispensable and increasingly important components of the system. Accurate simulations of the power electronic equipment dynamics become more critical for the equipment design and system construction, and are attracting increasing attention. 

Existing numerical simulation methods for power electronic equipment can be divided into two categories: (1) circuit-based methods where power semiconductor devices are modeled by circuit components \cite{TPE1998DeviceModels}; and (2) detailed multiphysics analysis methods where power semiconductor devices are modeled using PDEs, emphasizing on device-level physics\cite{Selberherr1984Analysis}. 

Circuit-based simulations are widely used for large-scale system prototyping \cite{TPE1998DeviceModels}. Depending on the various applications and the acceptable accuracy, different levels of simplification of the devices are adopted. As listed in Table \ref{tab:DeviceModelLevels}, the level-0 and level-1 models are commonly used for large-scale system simulations, and extensive studies have been conducted. Zhu \textit{et al.} and Lin \textit{et al.} explored approaches to device model segmentation and simplification to enhance computing efficiency \cite{Yicheng2019DSED,Dinavahi2017}. Furthermore, system- or equipment-level partitioning was analyzed by Jin \textit{et al.} and Yu \textit{et al.} to expand the simulation scales \cite{Jianfeng2023MAHE,Yu2023AutomaticPartitioning}. Moreover, various parallelization techniques have been investigated to boost the computational speed\cite{Jianfeng2023RegionFolding,TPE2018IGBT_GPU,ParallelComputing2023TIE}. These circuit-based simulators achieve rapid speed and satisfactory accuracy for a wide range of applications. 

However, the simplifications inherent in circuit based simulation methodologies limit their capacity to account for device dynamics under extreme conditions. This limitation is particularly pronounced for optimizing equipment performance, analyzing equipment failures, and determining the device's safe operating area (SOA) \cite{PowerDeviceLevels1995}---all of which are essential for reliable equipment design and failure analysis.

Therefore, simulations with PDE-modeled devices are indispensable. These simulations employ the level-3 model listed in Table \ref{tab:DeviceModelLevels}, and are carried out mainly using commercial technology computer-aided design (TCAD) simulators \cite{Sentaurus}. For better equipment design, Xu \textit{et al.} and Zhang \textit{et al.} examined the high-surge behaviors of devices in converters \cite{Xu2020Novel} and di/dt rates in DC breakers \cite{MixtureSolidStateSwitch2020TIE} using TCAD simulators, respectively. Similarly, for in-depth equipment analysis with the carrier characteristics inside the devices, Ren \textit{et al.} \cite{Ren2023Abnormal}, Stamenkovic \textit{et al.} \cite{APEC2019SoftSwitching} and Wang \textit{et al.} \cite{Wang2023Refinement} explored intricate device failure mechanisms and operational losses using realistic circuit typologies with TCAD tools. Moreover, to optimize device for equipment performances at realistic extreme conditions, Liu \textit{et al.} \cite{Liu2020UltraLow} and Ren \textit{et al.} \cite{Ren2023Optimal} conducted simulations for DC breakers and novel converters with TCAD tools. These PDE-based TCAD simulations are the only way to realize fine-grained device-level analysis and conduct DoEs, capturing the complicated carrier behaviors that circuit-based simulations often overlook \cite{Ren2022Deciphering,Ma2020Future}.

However, limited by the computational efficiency of existing commercial TCAD software, which were primarily designed for single-device simulations, the above studies were limited to 1 or 2 devices \cite{Ren2023Abnormal,ISPSD2020TwoCellCoSimulation}. Consequently, they could not effectively address more complex configurations such as power converters involving multiple devices in series or parallel configurations \cite{Zhang2018ModularDeviceSeries, Xu2023_HLCC_design}.

\begin{table*}[!ht]
\caption{Levels of semiconductor device models for power electronics simulations \label{tab:DeviceModelLevels}}
\centering
\begin{tabular}{|c|p{3.2cm}|p{2.5cm}|p{5.5cm}|p{4.3cm}|}
\hline
Level & Characteristics & Mathematical model & Typical applications & Typical simulators\\
\hline
0 & Ideal switch (switching between different resistances) & algebraic equation & Large scale power electronic system (PES) simulations with numerous switching cycles, where device-level switching dynamics are ignored. & Commercial PES simulators, i.e., PSCAD/EMTDC \cite{Xu2012PSCAD} and PLECS.
\\
\hline
1 & Simplified model built by circuit components& differential algebraic equations (DAEs) & General PES analysis with approximate device-level accuracy within the device SOA. & SPICE families, circuit simulators with customized device models \cite{Hefner1994Saber}.
\\
\hline
2 & Compact model with simplified physics& DAEs and simplified partial differential equations (PDEs) & General PES simulations with better device-level accuracy within SOA, like snubber and driver circuit design, and device power loss prediction.  & In-house codes with sophisticated compact models\cite{Lyu2018Compact}. \\
\hline
3 & Multi-physics model & PDEs & Device-level simulations with emphasis on accurate performance predictions of the device. & Commercial multi-physics TCAD solvers, e.g., Sentaurus TCAD.
\\
\hline
\end{tabular}
\end{table*}

Therefore, there exists a significant gap between the two existing methodologies and the requirements of power electronic equipment analysis, particularly for those DoEs related to device SOA. Circuit-based methods are efficient but lack detailed device behaviors; however, PDE-modeled simulations comprehensively account for the carrier-level device physics and are computationally inefficient.

\begin{figure}[!th]
\centering
\includegraphics[width=0.48 \textwidth]{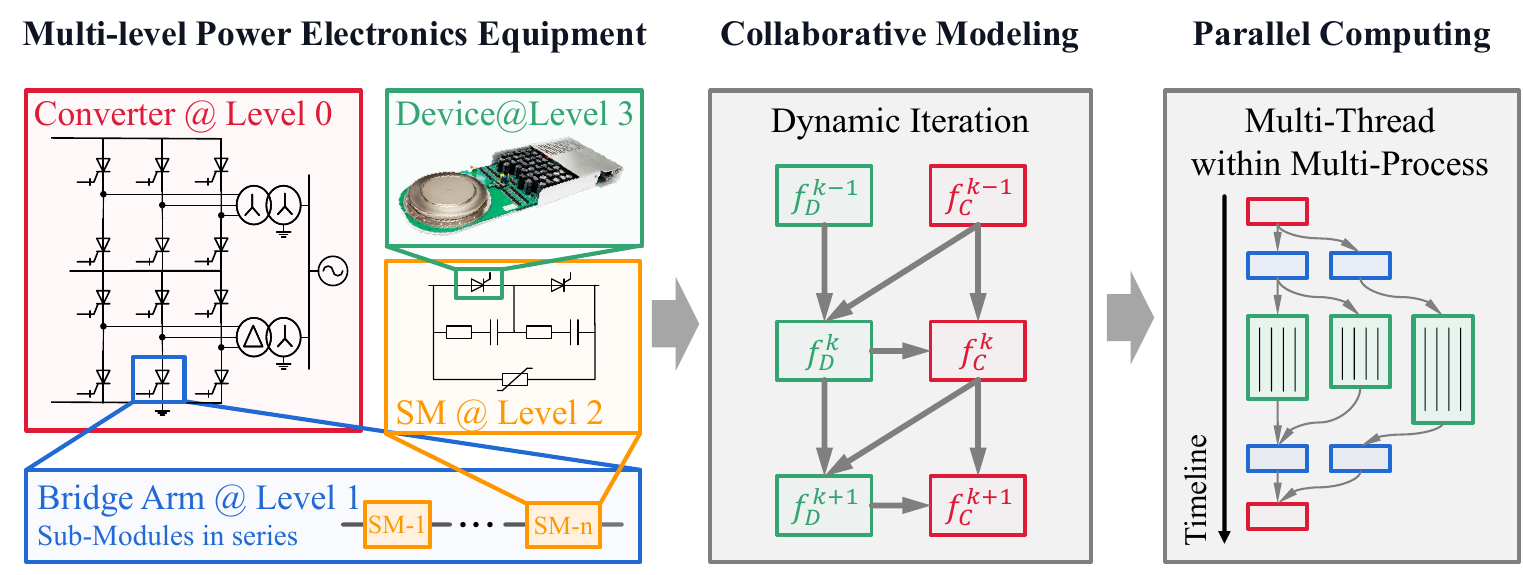}
\vspace{-5pt}
\caption{Overview of the proposed HPC framework.}
\label{fig:HHPOverview}
\end{figure}

To bridge this gap, this paper proposes a Hybrid-Parallel Collaborative (HPC) framework to analyze the power electronic equipment, which is modeled by PDAE systems where the PDEs are coupled with differential-algebraic equations (DAEs). The HPC framework is tailored to facilitate efficient collaborative simulations in a balanced manner and highlights (cf. Fig. \ref{fig:HHPOverview}):

1) \textbf{Integration of device physics with circuit dynamics}: The device physics is modeled by PDEs and the circuit dynamics is modeled by DAEs.  

2) \textbf{Efficient parallel computing}: The hybrid parallelization paradigm on shared and distributed memory with physics-based system partitioning boosts the dynamic iteration for complex power electronic equipment.

We emphasize that scenarios where circuit simulations suffice are not within the scope of our work. The objective of this work is to bridge the gap between high-fidelity device-level analysis and equipment-level simulations, integrating carrier-level precision into equipment-level studies while achieving high computational efficiency. This is especially beneficial in contexts that require an understanding of how microscopic processes, such as carrier dynamics, impact the macroscopic behavior of equipment in real-world applications \cite{Ren2023Abnormal, Liu2024Investigation, ABB2020LowCurrent, JB2024_IGBT}.

The remainder of this paper is organized as follows. Section II outlines the PDAE-modeling for power electronic equipment. Section III introduces the hybrid-parallel computing method with physics-based system partitioning for dynamic iteration. Section IV presents three detailed experiments derived from real-world engineering practices. We used the HPC framework to support a converter design workflow, including device optimization, converter optimization, and device-related failure analysis. Comparisons are made between the HPC framework, commercial softwares (Sentaurus TCAD, PSCAD and Simulink) and the measured experimental waveforms.

\section{Multiscale Modeling of Power Electronic Equipment using PDAE}

Power electronic equipment is a circuit system that incorporates power semiconductor devices, encompassing characteristics across multiple scales of interest depending on different applications and scenarios—from carrier-level processes to circuit module transients. To effectively capture these characteristics at each level, this section presents a modeling strategy integrating the device physics and circuit dynamics, which finally forms a coupled PDAE system. 

\subsection{PDE Modeling of Semiconductor Devices in Equipment}
The basic model for semiconductor devices is the drift diffusion model, where the electrostatic potential $\psi$ and carrier densities $n,p$ inside a device are described by the following PDEs with properly assigned boundary conditions \cite{Selberherr1984Analysis}:
\begin{equation}
    \label{equ:TCADGoverningEquations}
    \left\{
    \begin{aligned}
        &0 = \nabla^2 \psi+\frac{q}{\varepsilon}\left(p-n+N_D^{+}-N_A^{-}\right) \\
        &0 = \frac{\partial n}{\partial t} - \nabla \cdot \left(\mu_n n E_n+\mu_n V_T \nabla n\right) + R_{n} - G_{n}\\
        &0 = \frac{\partial p}{\partial t} + \nabla \cdot \left(\mu_p p E_p-\mu_p V_T \nabla p\right) + R_{p} - G_{p}
    \end{aligned}
    \right.
\end{equation} 
where $q$ is the elementary charge and $\varepsilon$ is the dielectric constant of semiconductor; $N_D^{+}, N_A^{-}$ denote ionized dopant concentrations; $E_n, E_p$ are the effective electric field considering energy bandgap narrowing effect; $\mu_n,\mu_p$ denote the mobility of the carriers given by Philips unified mobility model; $V_T$ denotes thermal voltage; $R_n,R_p$ denote recombination rates of electrons and holes by the Shockley-Read-Hall recombination and Auger recombination effects, respectively; and $G_n,G_p$ denote generation rates of electrons and holes by impact ionization, respectively. Every physical model is nonlinear and incorporates complicated physical mechanisms like Fermi-Dirac distribution of carriers. For detailed explanations of these models and their parameters, the interested readers may refer to \cite{sze2021physics,Selberherr1984Analysis}.

Simulating the microscopic characteristics of semiconductor devices is a significant challenge because Eq. \eqref{equ:TCADGoverningEquations} is not only strongly coupled, but also highly convection dominated, and traditional algorithms often suffer from the numerical diffusion and numerical oscillation \cite{amc,jcp1, jcp2}.

\begin{figure}[!ht]
\centering
\includegraphics[width=0.35\textwidth]{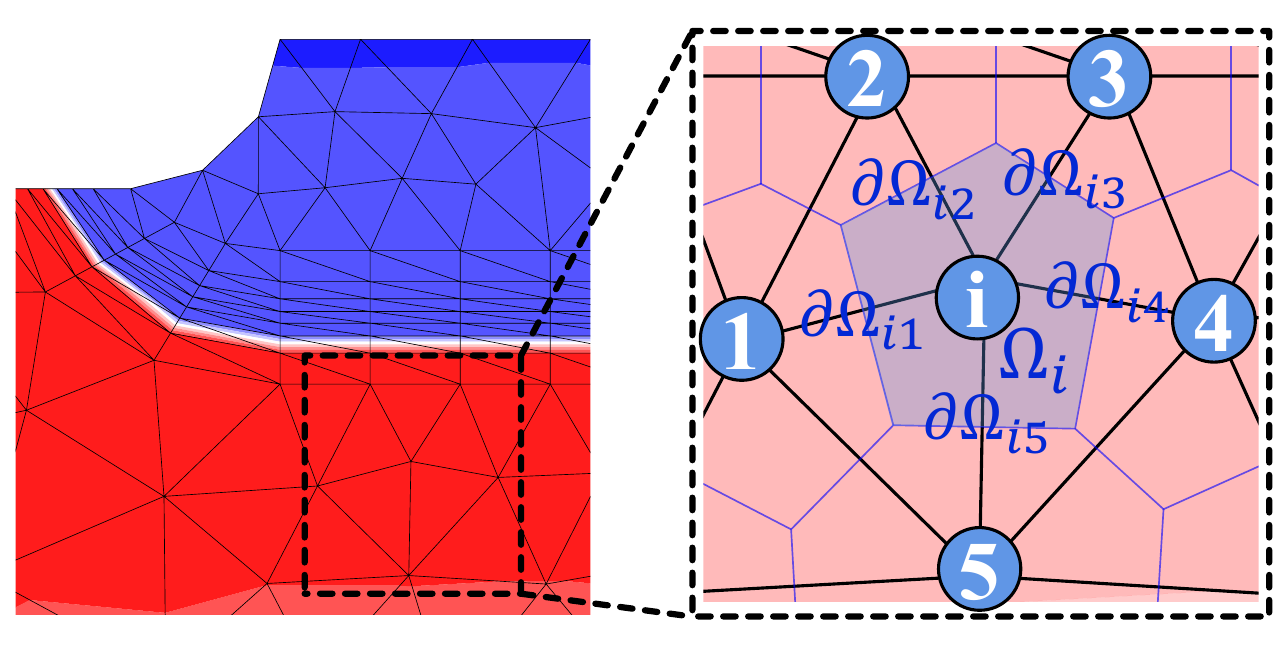}
\caption{Finite volume Scharfetter-Gummel discretization for the device model, where the mesh is simplified or coarsened only for illustration purpose.}
\label{fig:SG_Mesh}
\end{figure}

The HPC framework realizes the numerical solution to Eq. \eqref{equ:TCADGoverningEquations} by the finite volume Scharfetter-Gummel scheme \cite{Selberherr1984Analysis}, which discretizes the nonlinear convection-dominated PDEs on every mesh vertex as depicted in Fig. \ref{fig:SG_Mesh}, where the black lines are mesh edges and blue lines are the interfaces of control volumes. For vertex $i$ and the associated control volume $\Omega_i$, the discretized nonlinear governing equations are
\begin{equation}
    \label{equ:SG_discretization}
    \left\{
    \begin{aligned}
        0 & =\sum_{j\in\text{N}(i)}\int_{\partial \Omega_{ij}}\frac{\psi_j-\psi_i}{e_{ij}}\mbox{d}S +\\
        &\int_{\Omega_i} \frac{q}{\varepsilon}\left(p_i-n_i+N_{D,i}-N_{A,i}\right) \mbox{d}\Omega \\
        0 & = \int_{\Omega_i}\left(\frac{\mbox{d} n_i}{\mbox{d} t}+R_i-G_i\right) \mbox{d}\Omega+ \\
        &\sum_{j\in\text{N}(i)}\int_{\partial \Omega_{ij}}\frac{D_n}{e_{ij}}\left(n_i\text{B}\left(\frac{\psi_i-\psi_j}{V_T}\right)-n_j\text{B}\left(\frac{\psi_j-\psi_i}{V_T}\right)\right)\mbox{d}S \\
        0 & = \int_{\Omega_i}\left(\frac{\mbox{d} p_i}{\mbox{d} t}+R_i-G_i\right) \mbox{d}\Omega+ \\
        &\sum_{j\in\text{N}(i)}\int_{\partial \Omega_{ij}}\frac{D_p}{e_{ij}}\left(p_i\text{B}\left(\frac{\psi_j-\psi_i}{V_T}\right)-p_j\text{B}\left(\frac{\psi_i-\psi_j}{V_T}\right)\right)\mbox{d}S
    \end{aligned}
    \right.
\end{equation} 
\noindent
where $\text{N}(i)$ is the collection of neighbors of vertex $i$; the diffusion coefficients $D_n=\mu_nV_T$, $D_p=\mu_pV_T$; $\text{B}(x)=\frac{x}{\exp(x) - 1}$ is the Bernoulli function; $e_{ij}$ is the length of edge $ij$ and $\partial \Omega_{ij}$ is the facet of $\Omega_i$. This discretization yields a large nonlinear system whose dimensions are three times the number of vertices, and is then solved using Newton's method. Details on these numerical techniques can be found in  \cite{Leveque2002FiniteVolume,Brenner2008mathematicalFEM,Selberherr1984Analysis}.

At the electrodes of PDE-modeled devices, the device currents will interact with the circuit components of the equipment. These currents are determined by both the device internal variables and the electrode voltages. Typically, the currents flowing through electrode $i$ is given by
\begin{eqnarray}
    \label{equ:current}
    I^e_{i}(V^e, \psi, n, p) = \int_{\partial S_i}\Big(\varepsilon \frac{\partial E}{\partial t} + q\mu_n n E_n+\mu_n kT \nabla n  \nonumber \\
    + q\mu_p p E_p - \mu_p kT \nabla p \Big) \mbox{d}S, 
\end{eqnarray}
\noindent
where $E=-\nabla \phi$; $S_i$ is the surface area of the electrode $i$ and $V^e$ is the electrode voltage that serve as boundary condition; the superscript $e$ indicates that these current and voltage values are associated with an electrode that links the device and circuit. 

\subsection{Coupled PDAE System of Equipment and Its Decoupling through Dynamic Iteration}

Apart from the PDE-modeled devices, the remaining parts of the equipment can be abstracted by a circuit system governed by Kirchhoff's law, which can then be solved by the modified nodal analysis, resulting in a series of algebraic and ordinary differential equations, namely DAEs \cite{MNA1975}.

Therefore, the HPC framework abstracts the overall coupled system describing the multiscale dynamics of power electronic equipment as a nonlinear PDAE system \cite{SIAM2018PDAE}

\begin{equation}
    \label{equ:circuit_DAE}
    \begin{aligned}
    &f_D(\psi, n, p, V, I) = 0,\\
    &f_C(\psi, n, p, V, I) = 0,
    \end{aligned}
\end{equation}
\noindent
where $f_D, f_C$ denote the discretized device PDEs (Eq. \eqref{equ:SG_discretization}) and circuit DAEs, respectively. $V,I$ are the nodal voltages and branch currents in the circuit. The variables $\psi, n, p, V, I$ appear in both the device and circuit systems, reflecting their close coupling. Specifically, this interconnection is facilitated by the coupling through device electrode voltages and currents, as detailed in Eq. \eqref{equ:current}.

Coupled solution of this system presents significant computational challenges due to the huge difference in magnitude of the solution variables. These variables range from approximately $0 \sim 10^4$ V in circuit DAEs to around $\sim 10^{21} \rm cm^{3}$ in device PDEs. This multiscale variability in solution variables not only complicates the selection of initial guesses for Newton's iteration, but also results in extremely high condition number of the Jacobian matrix.

Therefore, the HPC framework incorporates the dynamic iteration \cite{SIAM2018PDAE} to tackle this strong and highly nonlinear coupling, handling the PDEs and DAEs by a decoupled iterative approach via the boundary condition exchange. Such decoupling helps to mitigate the sensitivity of the Newton iteration to the initial guess. The iterations can be performed by Jacobi or Gauss-Seidel types. Our experiments show that Gauss-Seidel type iterations have much better convergence property in practices, thus we only adopt the Gauss-Seidel type iteration, which is given as follows:
\begin{subequations}
    \label{equ:coupled_DAE_GS}
    \begin{align}
    \text{Subsystem 1:} \ &f_D^{k+1}(\psi^{*}, n^{*}, p^{*}, V^{k} , I^{k}) = 0, \label{eq:GS_sub1} \\
    \text{Subsystem 2:} \ &f_C^{k+1}(\psi^{*}, n^{*}, p^{*}, V^{k+1}, I^{k+1}) = 0, \label{eq:GS_sub2}
    \end{align}
\end{subequations}
where $k$ denotes the iteration step and the variables with a superscript $*$ denote the intermediate device solution between iteration $k$ and $k+1$, which are then plugged into $f^{k+1}_C$ for circuit.

\begin{figure}[!ht]
\centering
\includegraphics[width=0.32 \textwidth]{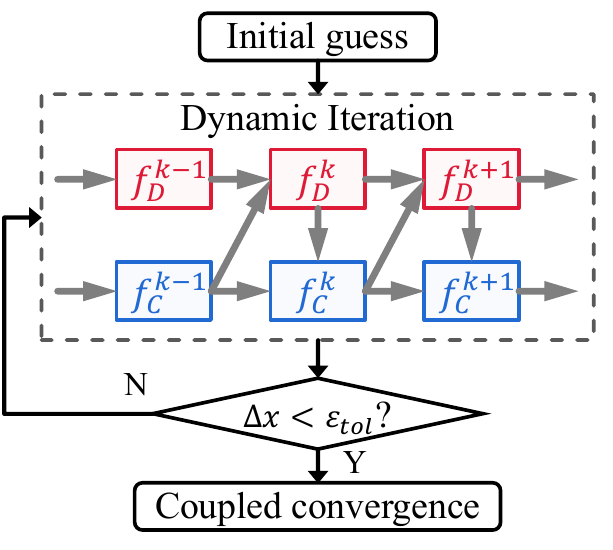}
    \vspace{-10pt}
\caption{Flowchart of the Gauss-Seidel type iteration.}
\label{fig:Jacobi_Gauss_Seidel}
\end{figure}

Fig. \ref{fig:Jacobi_Gauss_Seidel} presents the flowchart of the dynamic iteration, where the subscripts $C$ and $D$ denote the solution of nonlinear systems for the circuit and device subsystems, respectively. The gray arrows indicate the successive exchange of the coupling conditions.

For transient simulations, the implicit second-order backward differentiation formula (BDF-2) is used to discretize the time derivatives in both PDEs and DAEs, with an adaptive time-step control based on the local truncation error. More details on adaptive time stepping can be found in \cite{Selberherr1984Analysis}.

\subsection{Circuit Perspective for PDAE Systems Decoupled through Dynamic Iteration}

Dynamic iteration decouples the PDAE system. However, as shown in Eq. \eqref{equ:coupled_DAE_GS}, subsystem 2 (circuit DAE system) still relies on subsystem 1 (device PDE system). Therefore, when solving subsystem 2 using Newton's method, it is essential to incorporate the derivative of subsystem 2's equation with respect to subsystem 1's variables (i.e., $\psi, n, p$). Specifically, the term $\frac{\partial f_C}{\partial (\psi,n,p)}\frac{\partial (\psi,n,p)}{\partial (V,I)}$ must be included in the Jacobian matrix of subsystem 2.

As shown in Eq. \eqref{equ:current}, the two subsystems is only coupled by the devices' boundary voltages and currents $V^e, I^e$, therefore the Jacobian $\frac{\partial f_C}{\partial (\psi,n,p)}\frac{\partial (\psi,n,p)}{\partial (V,I)}$ is very sparse, and only the non-zero entries are required, namely
\begin{equation}
    \label{equ:NortonEquivalent}
    \begin{aligned}
    \left(\frac{\partial I^{e}_i}{\partial \mathbf{x}_D}\right)^T\frac{\partial \mathbf{x}_D}{\partial V^{e}_{j}},\\
    \end{aligned}
\end{equation}

\noindent
where $i,j$ are device electrodes and $\mathbf{x}_D$ represents the vector of internal variables of the device PDE, i.e., $\psi,n,p$ in \eqref{equ:TCADGoverningEquations}. The derivative $\frac{\partial I^e_i}{\partial \mathbf{x}_D}$ can be calculated analytically from Eq. \eqref{equ:current}; and the second term $\frac{\partial \mathbf{x}_D}{\partial V^{e}_{j}}$ can be derived from Eq. \eqref{eq:GS_sub1} by $\frac{\partial f_D\left(\mathbf{x}_D\left(V^e_j\right),V,I\right)}{\partial V^e_j} = 0$, i.e.,
\begin{equation}
    \label{equ:NortonEquivalentDxDv}
    \begin{aligned}
    J_D\frac{\partial \mathbf{x}_D}{\partial V^e_{j}}=-\frac{\partial f_D}{\partial V^e_{j}},
    \end{aligned}
\end{equation}
where $J_D$ are the Jacobian matrix of the discretized device PDEs (Eq. \eqref{equ:SG_discretization}).

We observe that the Jacobian shown in Eq. \eqref{equ:NortonEquivalent} possesses the physical unit as a conductance. Consequently, we define it as an equivalent conductance by
\begin{equation}
    \label{equ:EquivalentConductance}
    \begin{aligned}
    G_{ij}=\left(\frac{\partial I^{e}_i}{\partial \mathbf{x}_D}\right)^T\frac{\partial \mathbf{x}_D}{\partial V^{e}_{j}}
    =-\left(\frac{\partial I^{e}_i}{\partial \mathbf{x}_D}\right)^T\left(J_D\right)^{-1}\frac{\partial f_D}{\partial V^e_{j}}.\\
    \end{aligned}
\end{equation}

\begin{figure}[!ht]
\centering
\includegraphics[width=0.5 \textwidth]{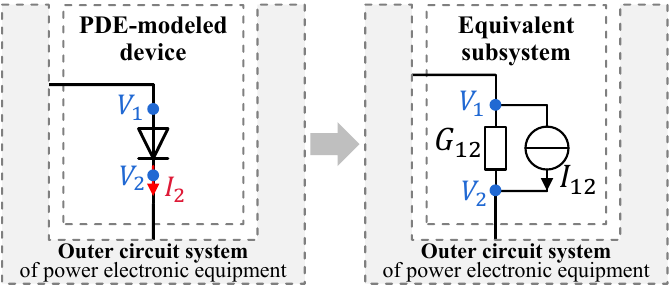}
\caption{Circuit perspective for Gauss-Siedel type dynamic iteration. Left: PDE-modeled device. Right: Equivalent subsystem derived from implicit derivative.}
\label{fig:DeviceSystemCoupling}
\end{figure}

Therefore, from a circuit perspective, the coupling between the PDE-modeled device and the remaining circuit is modeled by equivalent conductance $G_{ij}$ and parallel-connected current sources. Taking Fig. \ref{fig:DeviceSystemCoupling} for example, the current source $I_{12}$ is the remainder of the current through the PDE-modeled device ($I_2$) and its associated first-order approximation, namely
\begin{equation}
    \label{equ:EquivalentCurrentSource}
     I_{12}= I_2 - G_{12}\left(V_1-V_2\right).
\end{equation}

In this way, the Gauss-Seidel type dynamic iteration serves as a pivotal technique integrating the device-physics and the circuit, and avoids the coupled solution of unknowns with significant differences in magnitude. It should be noted that although the dynamic iteration decouples the PDAE system, no additional error is introduced because the whole system is solved iteratively until the convergence is achieved.
\begin{figure}[!ht]
\centering
\includegraphics[width=0.45 \textwidth]{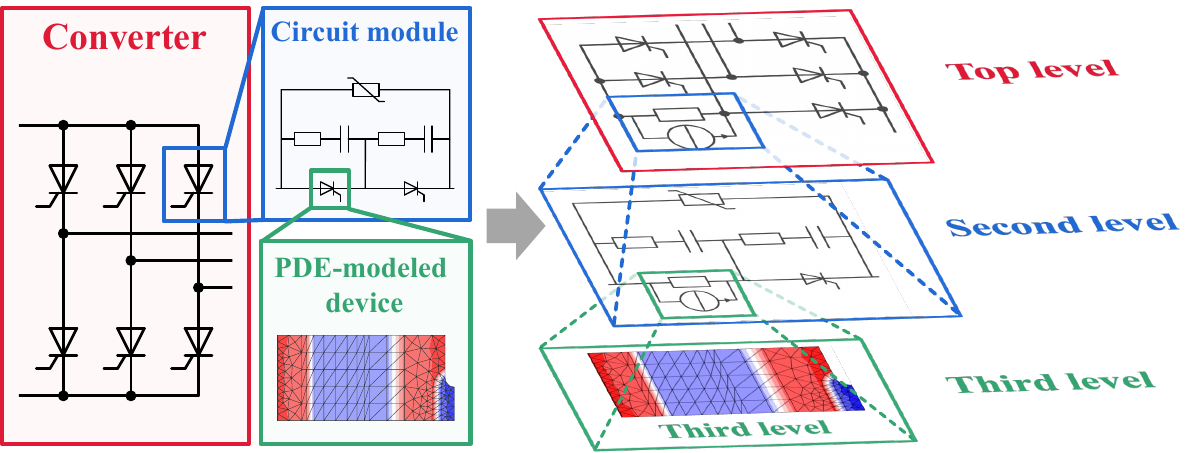}
\caption{A simplified illustration of a converter with 3 levels of equivalent subsystems from a circuit perspective.}
\label{fig:Converter3Level}
\end{figure}

It is worth noting that the chain rule in Eq. \eqref{equ:NortonEquivalent} can be applied to both PDE and DAE systems within the HPC framework. This facilitates universal modeling across various levels of devices and circuit modules in power electronic equipment, thereby enhancing the flexibility of solving the multiscale PDAE system. The modeling of a converter with three-level hierarchies is demonstrated in Fig. \ref{fig:Converter3Level} as an example, where the circuit part is modeled into two levels and the device is the third, interconnected by the equivalent conductances and current sources derived from the derivatives of lower-level subsystems.

\section{Hybrid-Parallel Computing Method for Dynamic Iteration}

\label{section3_hybrid_parallel}

As manifested in Eq. \eqref{equ:circuit_DAE}, the PDAEs for circuits and devices are tightly coupled. Conventional methods that handle strongly coupled devices and circuits within a single system are hardly conducive to efficient parallel computing. However, as is evident from Eq. \eqref{equ:coupled_DAE_GS}, the dynamic iteration separates the PDAEs into PDEs and DAEs, thus facilitating the independent sequential solution of the circuit and device. Therefore, we introduce a hybrid-parallel computing method to boost the efficiency of the dynamic iteration. It incorporates physics-based system partitioning and a synchronization mechanism to minimize communication overhead and guarantee stability and convergence across consecutive time steps.

\begin{figure}[!ht]
\centering
\includegraphics[width=0.45\textwidth]{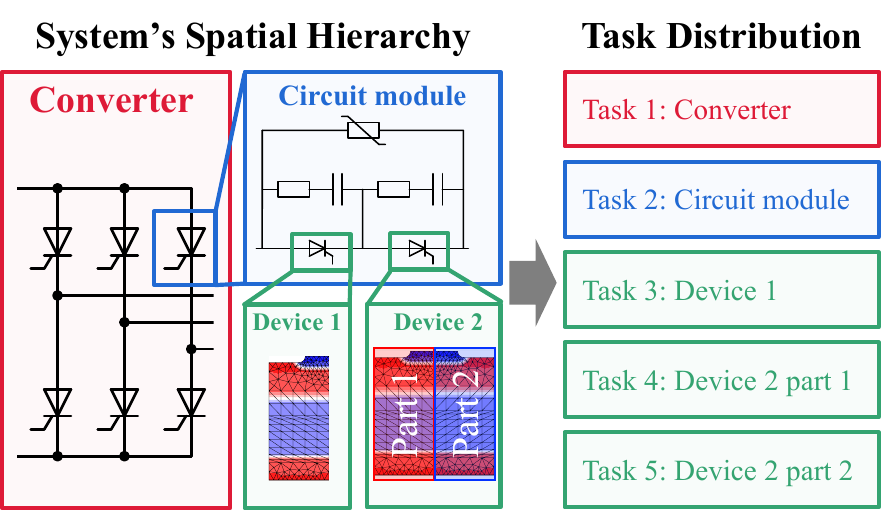}
\caption{Illustration of the proposed physics-based system partitioning.}
\label{fig:physics_task_distribution}
\end{figure}

To illustrate the strategy, we take a simplified converter in Fig. \ref{fig:physics_task_distribution} as an example. This system is initially divided into three types of subsystems: the converter bridge, series circuit modules, and PDE-modeled devices, based on the levels of physical coupling. Each pair of subsystems exhibits weak coupling. For larger PDE-modeled subsystems such as Device 2 in Fig. \ref{fig:physics_task_distribution}, further subdivision into two parts with negligible flux coupling can successively be utilized to split the domain. After the partition, different subsystems are assigned to different tasks. The partitions are not exclusive, and such physics-based partitioning reduces the communication overhead.

\begin{figure}[!ht]
\centering
\includegraphics[width=0.38\textwidth]{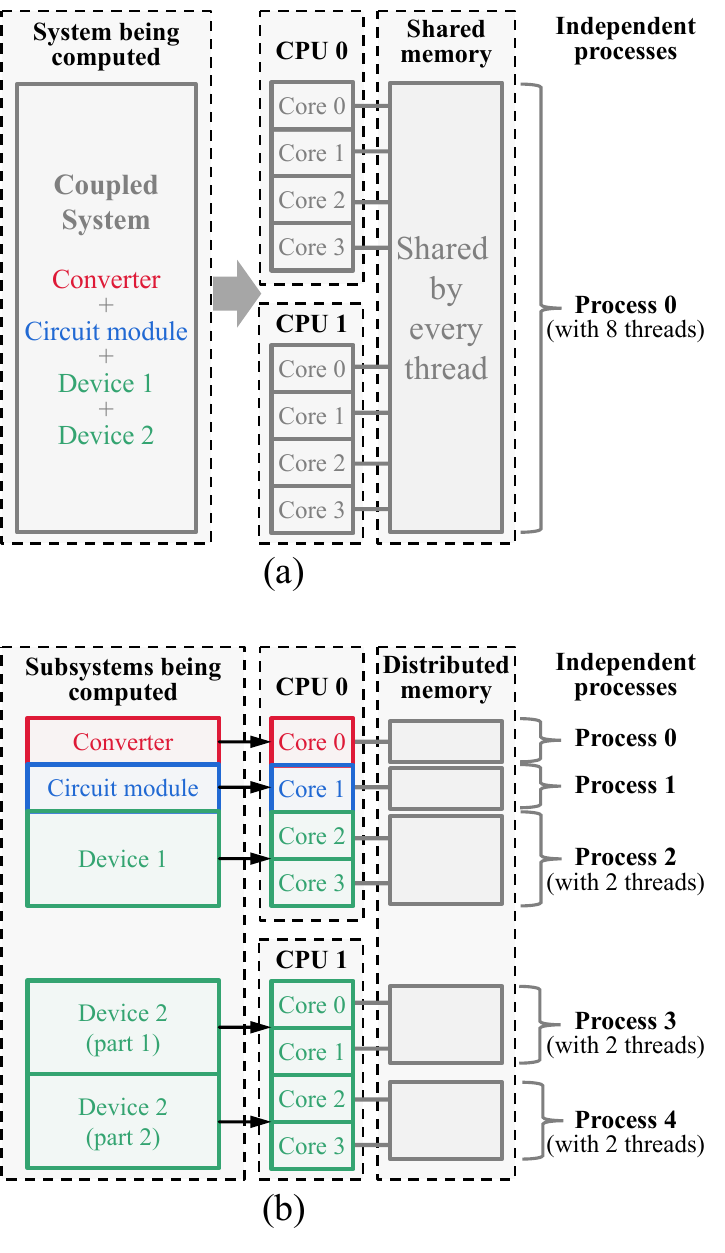}
\vspace{-15pt}
\caption{Illustration of conventional and proposed parallel computation approaches. (a) Conventional parallel scheme for circuit-device co-simulations. (b) Proposed hybrid-parallel scheme.}
\label{fig:parallel_scheme_compare}
\end{figure}

In addition to the partitioning, Fig. \ref{fig:parallel_scheme_compare} illustrates two different parallel approaches. In Fig. \ref{fig:parallel_scheme_compare}(a), conventional methods relying solely on multi-threading distribute the entire system across CPU cores, and the memory is shared by different threads. As the number of CPU cores (or threads) increases, the thread synchronization overhead increases disproportionately, leading to diminishing parallel efficiency and a plateau in speedup gains. In contrast, Fig. \ref{fig:parallel_scheme_compare}(b) demonstrates the hybrid-parallel framework, which assigns loosely coupled subsystems to separate processes on distributed memory while leveraging multi-threading within processes for tightly coupled tasks. This approach minimizes inter-process communication, enhances scalability across distributed-memory computing clusters, and maintains flexibility.

We wish to emphasize that dynamic iteration enables the decoupling and parallel resolution of these distinct tasks. Subsequently, the boundary conditions between various subsystems are exchanged iteratively to manage their interdependencies until a convergent solution is achieved. To guarantee stability and convergence, an appropriate synchronization strategy for different tasks is required.

\begin{figure}[!ht]
\centering
\includegraphics[width=0.44\textwidth]{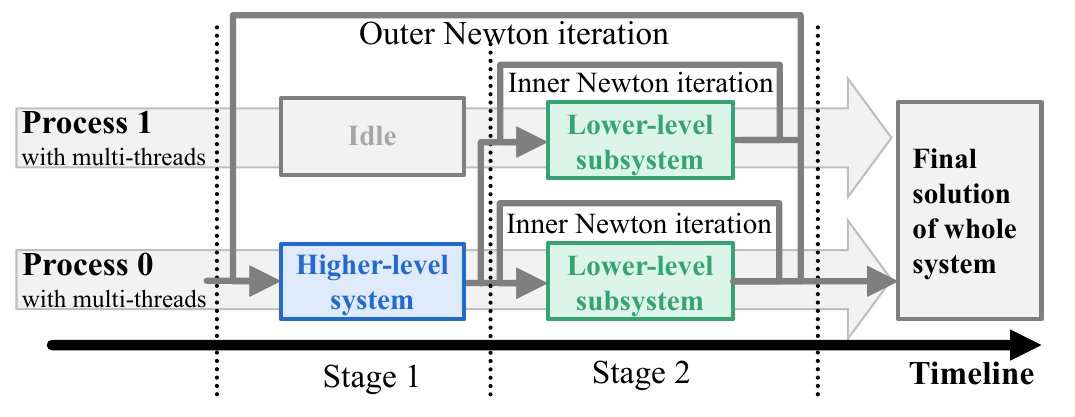}
                \vspace{-10pt}
\caption{The synchronization for Gauss-Seidel type dynamic iteration.}
\label{fig:HybridPallelComputing}
\end{figure}

Fig. \ref{fig:HybridPallelComputing} depicts the synchronization for the Gauss-Seidel type iteration. In Stage 1, the higher-level systems, specifically the circuit equations (Eq. \eqref{eq:GS_sub2} in our example), are solved. Subsequently, in Stage 2, the Newton iterations for the lower-level subsystems (illustrated by Eq. \eqref{eq:GS_sub1} in our example) and the equivalent circuit construction of the subsystems are carried out in parallel. Despite the inevitable nested recursion observed in Fig. \ref{fig:HybridPallelComputing}, this method can significantly reduce the overall computational time compared to solving a unified large system consisting of all equations. Quantitatively, the numerical complexity, which mainly comes from solving linear systems, reduces from $\sim O(n^3)$ to $\sim O\left(k(\frac{n}{k})^3\right)$ where $n$ is the number of equations in a larger system and $k$ is the number of partitioned subsystems of the devices.

It is worthy of mention that the most time-consuming part, specifically, solving the partial differential equation (PDE) systems of the devices, is effectively parallelized. As a result, the overall parallel efficiency is high. This allows the proposed framework to simulate power electronic equipment with multiple devices within a time frame comparable to that of simulating a single device.

\section{Experimental Validation and Case Studies}

\subsection{Overview of Experiments}
The proposed HPC framework aims to bridge the gap in analyzing how microscopic phenomena, like carrier dynamics, affect macroscopic equipment performance in real-world applications \cite{Liu2024Investigation, MixtureSolidStateSwitch2020TIE, ISPSD2020TwoCellCoSimulation}, and is designed for scenarios that require both device-level and circuit-level dynamics, such as device-equipment co-design, co-optimization, and failure analysis under extreme conditions.

Consequently, we carried out three sets of experiments to assess the precision and efficacy of the proposed HPC framework within our in-house solver. These experiments correspond to key stages of realistic converter development: device-level optimization, equipment-level design optimization, and analysis of failures induced by carrier dynamics.

\begin{figure*}[!th]
\centering
\includegraphics[width=\textwidth]{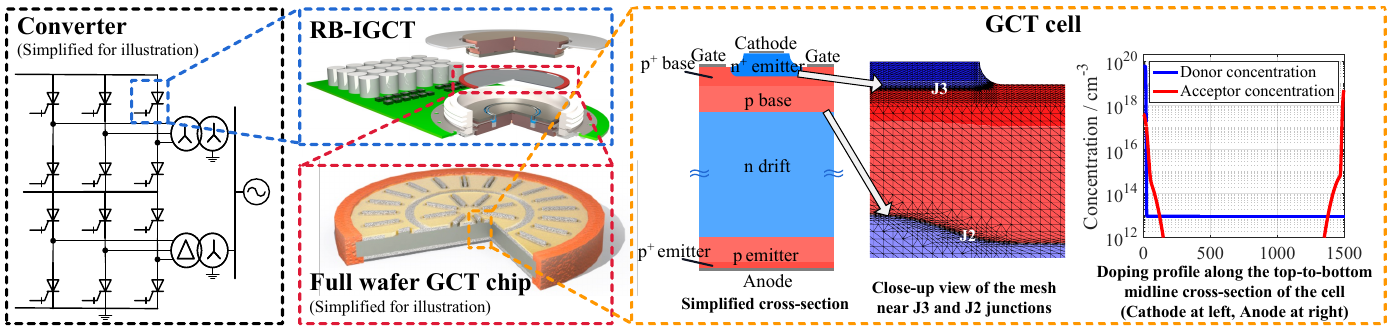}
\caption{Illustration of the H-LCC converter, RB-IGCT device, GCT wafer and one GCT cell, where red and blue areas represent P-type and N-type regions, respectively.}
\label{fig:IGCT_overview}
\end{figure*}

For comprehensive comparison and validation in realistic scenarios, the hybrid line commutated converter (H-LCC) \cite{Xu2022HLCC} is selected as an example topology. The H-LCC typically employs dozens to hundreds of power semiconductor devices in series to withstand high blocking voltages, making it a suitable benchmark to assess the scalability and accuracy of the proposed framework. The selected device is the reverse blocking integrated gate-commuted thyristor (RB-IGCT) (shown in Fig. \ref{fig:IGCT_overview}), which consists of thousands of parallel-connected cells with inherent variability among cells and devices. The RB-IGCT, known for its high current surge capability and high potential for high-voltage Direct Current (HVDC) applications \cite{Zeng2019IGCTMMC}, serves as an ideal representative for our case studies in device- and converter-level analysis.

It is noted that while the H-LCC topology is employed for validation purposes in this study, the proposed framework is versatile and applicable to a variety of other topologies, particularly those involving series and parallel configurations of multiple devices, thereby expanding its utility in the field of power electronics. Furthermore, the proposed HPC framework  is also adaptable for use with other semiconductor devices like IGBTs.

Numerical experiments were conducted on a Linux cluster with five nodes, each having 64 physical CPU cores and 256 GB RAM. We used the commercial software Synopsys TCAD Sentaurus Device, which is the state-of-the-art semiconductor device multi-physics solver.

\begin{figure}[!ht]
\centering
\includegraphics[width=0.49\textwidth]{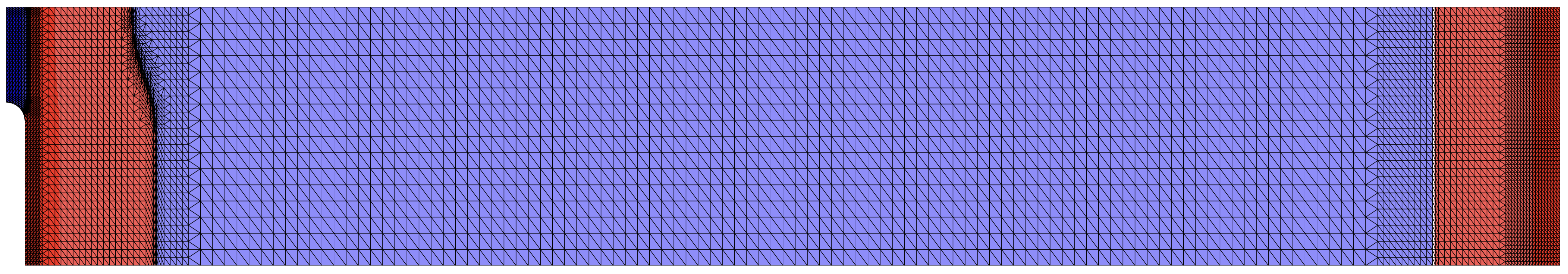}
\vspace{-10pt}
\caption{A mesh of an RB-IGCT cell used in the experiments.}
\label{fig:RB_cell_mesh_full}
\end{figure}

To ensure fair comparisons, we used the same settings as the commercial solvers. The error tolerance for Newton’s method was consistent, as specified in Table \ref{tab:Newton_parameters}. The maximum and minimum allowed time step sizes were 10 $\si{\micro s}$ and $10^{-6}$ $\si{\micro s}$, respectively. Throughout the experiments, the triangular meshes were used (cf. Fig. \ref{fig:RB_cell_mesh_full}): 16116 mesh cells for each of the 10 cells in Subsection IV-B, 18824 cells for each of all 160 devices in Subsection IV-C, and 16210 cells for GCT1 and 15597 cells for GCT2 in Subsection IV-D. The following physical models were incorporated: Fermi-Dirac carrier distribution, Philips mobility, high field mobility saturation, incomplete ionization, SRH recombination, Auger recombination, bandgap narrowing effect and impact ionization.

\begin{table}[!ht]
\renewcommand{\arraystretch}{1.5} 
\caption{Error Control Parameters of Newton's method\label{tab:Newton_parameters}}
\centering
\begin{tabular}{ccc}
\hline
\hline
Equation & Absolute tolerance & Relative tolerance \\
\hline
Poisson equation & $10^{-26} \rm C$ & $10^{-5}$\\
Electron continuity equation & $5\times 10^{-18} \rm A$ & $10^{-5}$\\
Hole continuity equation &  $5\times 10^{-18} \rm A$ & $10^{-5}$\\
Circuit equations &  $ 10^{-5}$ ($\rm V$ or $\rm A$) & $10^{-5}$\\
\hline
\hline
\end{tabular}
\end{table}

\begin{table}[!ht]
\renewcommand{\arraystretch}{1.5} 
\caption{Doping Parameters of the RB-IGCT\label{tab:IGCT_parameters}}
\centering
\begin{tabular}{ccc}
\hline
\hline
Doping region & (Peak) doping concentration / $\rm cm^{-3}$ & Depth / $\si{\micro \metre}$\\
\hline
n$^+$ emitter & $1.2\times 10^{20}$ & 23\\
p$^+$ base & $5\times 10^{17}$ & 60\\
p base & $1\times 10^{15}$ & 80\\
n drift& $9.8\times 10^{12}$ & 1250\\
p emitter & $1\times 10^{15}$ & 80\\
p$^+$ emitter & $5\times 10^{18}$ & 10\\
\hline
\hline
\end{tabular}
\end{table}

\subsection{Device Optimization With Parallel-Cell Simulations}
Optimizing the turn-OFF performance of customized power semiconductor devices is critical for a high-performance converter. The total turn-OFF current of an IGCT is much smaller than the sum of each cell (e.g., only 20\% of the sum) due to the non-uniform current distribution across the IGCT wafer. Therefore, balancing the current distribution is imperative for increasing maximum controllable current. This can be be effectively achieved through the lateral optimization of IGCTs \cite{ISPSD2014Fortified,Lyu_2018_Optimization}, such as modifying the carrier lifetime.

\begin{figure}[!ht]
\centering
\includegraphics[width=0.4\textwidth]{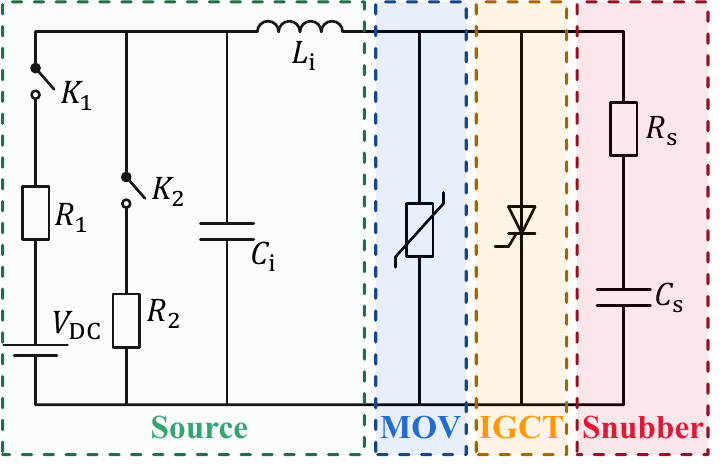}
\vspace{-10pt}
\caption{Circuit used for the lateral optimization of IGCT, based on a topology of H-LCC bridge arm in \cite{WZZ2024_VoltageEqualization}. Parameters can be found in \cite{WZZ2024_VoltageEqualization}.}
\label{fig:10RingTestCircuit}
\end{figure}

\begin{figure}[!ht]
\centering
\includegraphics[width=0.48\textwidth]{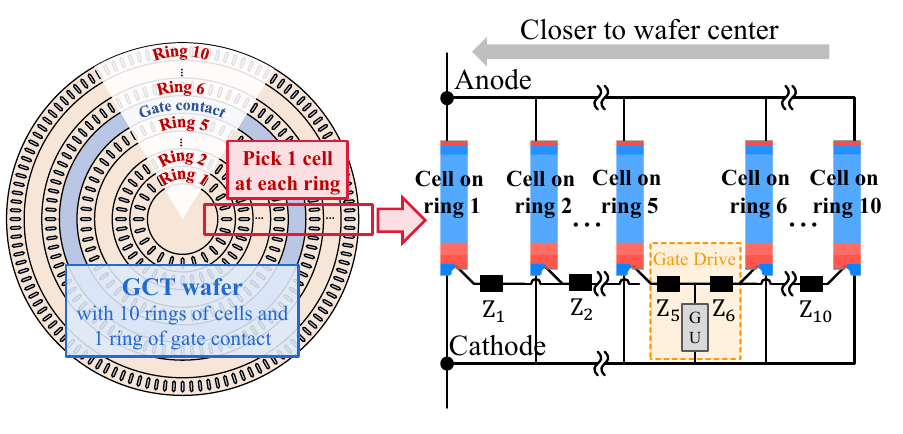}
\vspace{-10pt}
\caption{Full-wafer 10 ring structure and equivalent 10-cell model of the RB-IGCT considering stray parameters.}
\label{fig:10RingCircuit}
\end{figure}

In this study, we leverage the proposed HPC framework to enable collaborative simulations of a complete RB-IGCT wafer and a circuit shown in Fig. \ref{fig:10RingTestCircuit}. The full 10-ring structure of the IGCT wafer, illustrated in Fig. \ref{fig:IGCT_overview}, is modeled as a 10-cell circuit in Fig. \ref{fig:10RingCircuit}, where each cell represents one ring. Stray inductances and resistances, measured experimentally in \cite{Lyu_2018_Optimization}, are incorporated as circuit components, and simulated together with the driver circuit. We wish to remark that this circuit modeling approach used in this experiment, which includes the integration of stray parameters as circuit components, can also be applied to other power semiconductor devices, such as GaN and SiC switches.

The parameters to be optimized are the lifetimes of the carriers. The values of the initial setup are shown in Table \ref{tab:cell_lifetime_optimzation}. The optimization was carried out manually through a trial-and-error DoE approach. Both the HPC framework and commercial software were utilized for these simulations, with a maximum utilization of 80 CPU cores.

\begin{figure}[!ht]
\centering
\includegraphics[width=0.5\textwidth]{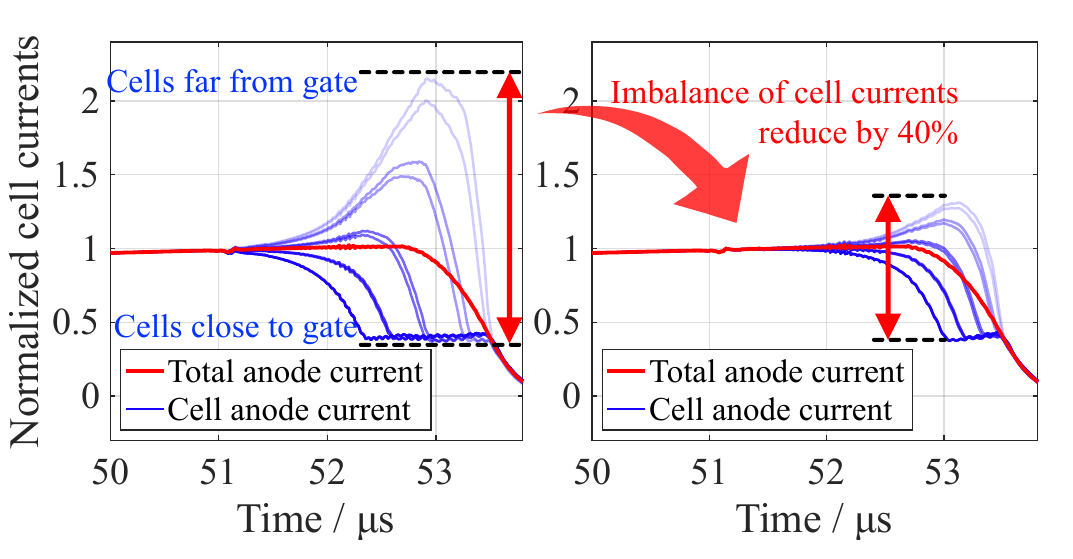}
\vspace{-12pt}
\caption{An illustrative example of full-wafer optimization by 10-cell PDAE modeling. Left: before optimization. Right: after optimization.}
\label{fig:10RingCurrentsBalance}
\end{figure}

By tuning the electron and hole lifetimes as in Table \ref{tab:cell_lifetime_optimzation}, the current imbalance is reduced by 40\%, as shown in Fig. \ref{fig:10RingCurrentsBalance}. This highlights the potential for optimized RB-IGCT designs with improved current control and reliability in converter applications.

\begin{table}[!ht]
\renewcommand{\arraystretch}{1.5} 

\caption{Optimization of carrier lifetime \label{tab:cell_lifetime_optimzation}}
\centering
\begin{tabular}{p{2cm}p{2.3cm}p{2.3cm}}
\hline
\hline
Cell ring number & Carrier lifetime\newline before optimization & Carrier lifetime\newline after optimization \\
\hline
Ring 3,4,5,6,7,8 & Electron: \SI{500}{\micro s} \newline Hole: \SI{150}{\micro s} & Electron: \SI{400}{\micro s} \newline Hole: \SI{120}{\micro s}\\
Ring 1,2,9,10 & Electron: \SI{500}{\micro s} \newline Hole: \SI{150}{\micro s} & Electron: \SI{320}{\micro s} \newline Hole: \SI{100}{\micro s}\\
\hline
\hline
\end{tabular}
\end{table}

\begin{figure}[!ht]
\centering
\includegraphics[width=0.485\textwidth]{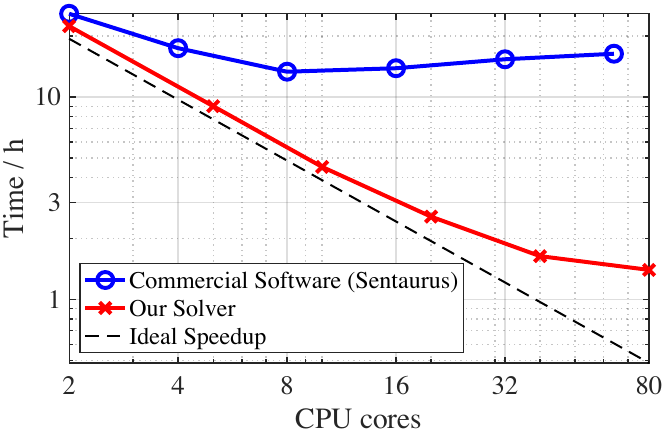}
\vspace{-8pt}
\caption{Comparison of the time for a single turn-OFF simulation for the 10-cell model in Fig. \ref{fig:10RingCircuit}.}
\label{fig:10RingSpeedCompare}
\end{figure}

It should be noted that during the trial-and-error process, repeated co-simulations were conducted.
Fig. \ref{fig:10RingSpeedCompare} shows the computation time of a single run. The performance of Sentaurus TCAD slightly degrades after 8 cores because its parallelization overhead exceeds the computational gains after 8 cores. Alternatively, the HPC framework demonstrated a 10-fold speed-up over the commercial TCAD software, reducing the solution time for a single turn-OFF simulation from 20 hours to less than 2 hours (Fig. \ref{fig:10RingSpeedCompare}). This acceleration significantly reduces the total time required for subsequent analysis and iterative optimizations, making it feasible to conduct extensive DoE studies for practical RB-IGCT designs.

The aim of this experiment is to demonstrate that our HPC framework can significantly accelerate real-world device-level research. Although, in practice, the carrier lifetime values obtained during manufacturing may deviate from the designed parameters due to manufacturing deviations, these multiphysics simulation results still provide valuable insights to guide optimization directions.

Moreover, the speed improvement achieved in this experiment is based on the modeling and computational methods introduced in Section III. We emphasize that the full PDAE system is solved without any simplification. Therefore, the precision of the results is not compromised.

\subsection{Converter Optimization for Devices in Series}

A high voltage converter consists of many devices connected in series. The nonuniformity of devices arising from manufacturing variations significantly impacts the voltage distribution imbalance between different devices in converters and must be considered for a robust design. These challenges require converter analysis with PDE-modeled devices.

\begin{figure}[!ht]
\centering
\includegraphics[width=0.45\textwidth]{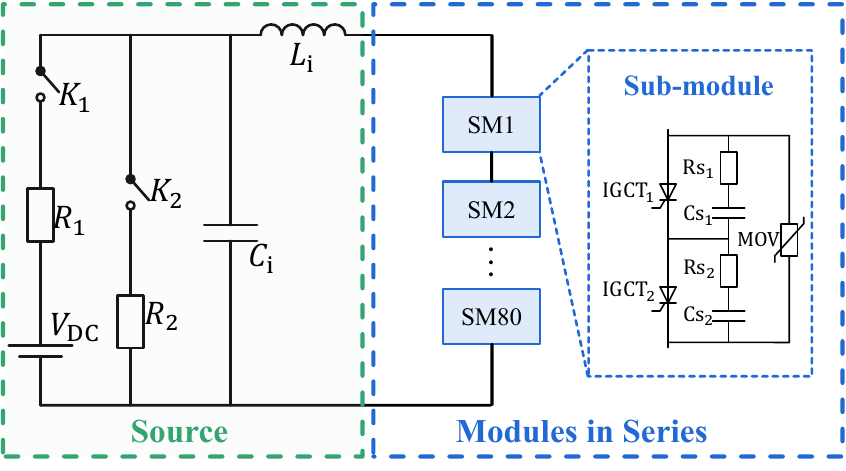}
\vspace{-12pt}      
\caption{Circuit for H-LCC bridge arm voltage equalization analysis with series modules. Parameters can be found in \cite{Xu2022HLCC}}
\label{fig:HLCCTopologyAndCompare}
\end{figure}

To demonstrate the capability of the HPC framework, an optimization of the voltage equalization circuit was conducted in an H-LCC \cite{Xu2022HLCC} bridge arm. The circuit configurations are based on prior H-LCC studies \cite{Xu2022HLCC, Xu2023_HLCC_design}, studying a bridge arm with 80 sub-modules connected in series. Each sub-module contains two RB-IGCTs, resulting in 160 devices in series conducting 5 kA of current during a simultaneous turn-OFF event. These devices incorporate 10\% random variations in doping profiles based on Table \ref{tab:IGCT_parameters}, generated using a uniform distribution, to reflect realistic manufacturing variability. This setup provides a representative scenario for analyzing voltage equalization in high-voltage converters.

To reduce the voltage imbalance among different devices and maintain an acceptable losses, the optimization focused on tuning the snubber capacitance $C_s$ \cite{Xu2022HLCC, Xu2023_HLCC_design}. The range for $C_s$ was chosen between $\SI{2.2}{\micro \farad}$ and $\SI{4.0}{\micro \farad}$ based on typical design guidelines for H-LCC converters; and repeatedly repeated trial-and-error computations were performed, balancing voltage uniformity with acceptable system losses and harmonic oscillations. Each simulation solved the nonlinear PDAEs with approximately 5 million DOFs per time step, marching through approximately 2600 time steps in total. 

\begin{figure}[!ht]
\centering
\includegraphics[width=0.48\textwidth]{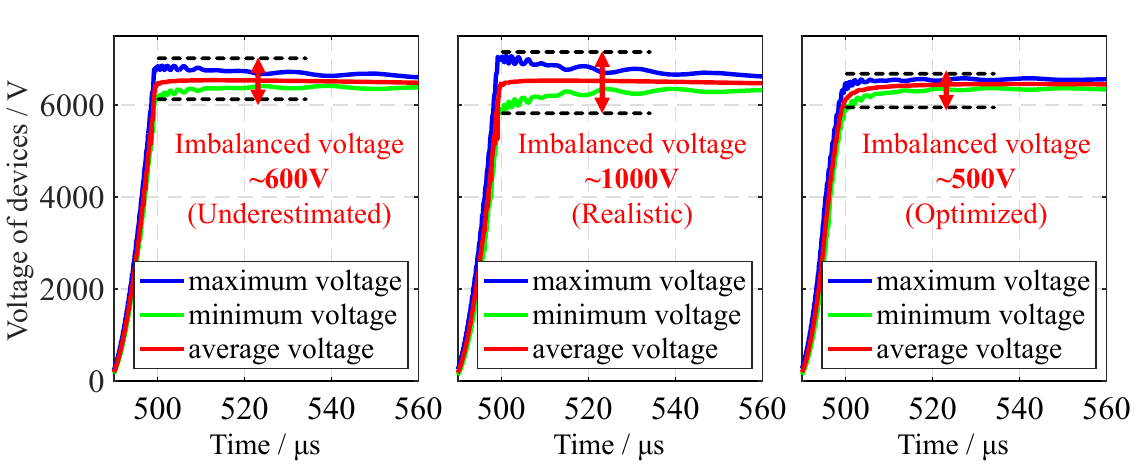}
\vspace{-5pt}
\caption{Voltage equalization analysis. Left: underestimated results simulated using 2 devices in series (only 1 sub-module in Fig. \ref{fig:HLCCTopologyAndCompare}); Middle: results simulated using 160 devices; Right: optimized results simulated using 160 devices.}
\label{fig:160SeriesVoltageImbalance}
\end{figure}

Fig. \ref{fig:160SeriesVoltageImbalance} further highlights the importance of full-size model simulations. A simplified model with only 2 devices underestimates the voltage imbalance by 40\%, emphasizing the need for detailed analysis of all 160 devices to achieve reliable results. By tuning the $C_s$ from $\SI{2.2}{\micro \farad}$ to $\SI{3.0}{\micro \farad}$, the voltage imbalance was significantly reduced. Although this result may not be globally optimal, it illustrates the capability of the HPC framework to efficiently support such optimization workflows.

\begin{figure}[!ht]
\centering
\includegraphics[width=0.485\textwidth]{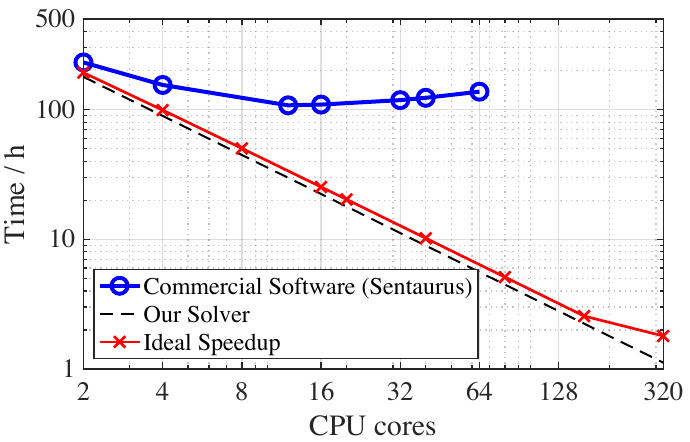}
\vspace{-6pt}
\caption{Comparison for the time of a single simulation for the 160-device-series model in Fig. \ref{fig:HLCCTopologyAndCompare}.}
\label{fig:160CellSpeedCompare}
\end{figure}

Fig. \ref{fig:160CellSpeedCompare} shows the runtime for a single simulation, where the HPC framework achieves a 60-fold speedup over commercial software, reducing the time from more than 5 days to less than 2 hours. This acceleration demonstrates the framework's scalability to leverage much more CPU cores with hybrid parallelization, which is shown in Fig. \ref{fig:10RingSpeedCompare} and Fig. \ref{fig:160CellSpeedCompare}.

We remark that this experiment's primary focus is not only on the optimization result but also on demonstrating the HPC framework's ability to handle large-scale simulations efficiently. By reducing simulation times from several days to hours, the framework enables workflows that were previously infeasible.

\subsection{Device Failure Analysis}
\begin{figure}[!htp]
\centering
\includegraphics[width=0.485\textwidth]{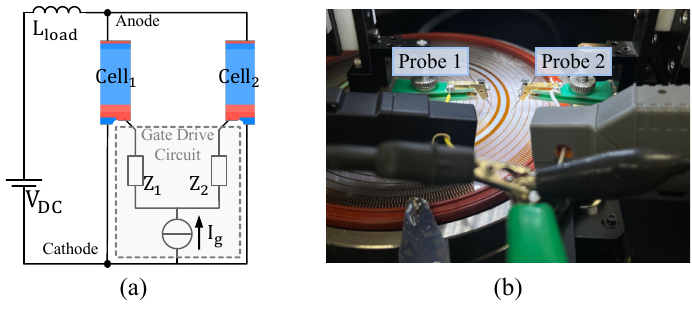}
\vspace{-12pt}
\caption{Experiment setup for abnormal device turn-ON. (a) Circuit topology. (b) Test platform.}
\label{fig:TwoCellTestCircuit}
\end{figure}

The above experiments have demonstrated the HPC framework's superior speed compared to commercial TCAD software. Moreover, it can serve as a practical tool for device failure analysis, providing more internal physical insights. An important application is the analysis of power converter abnormalities caused by semiconductor devices. Reference \cite{Ren2023Abnormal} reported an abnormal low voltage turn-ON phenomenon caused by non-identical dynamic characteristics between cells of an RB-IGCT and a test platform was established to measure the transient waves (cf. Fig. \ref{fig:TwoCellTestCircuit}). 

To demonstrate the capabilities and potential applications of the HPC framework in such realistic scenarios, we replicated the cell-level experiment and simulation in \cite{Ren2023Abnormal} with the same settings. The two cells are with identical structure and doping parameters in Table \ref{tab:IGCT_parameters} except for the p$^+$-base peak doping concentration in GCT1, which was set to $5.5\times 10^{17} \rm cm^{-3}$, i.e., 10\% higher than the design value. This variation reflects the typical deviations from ion implantation encountered in IGCT manufacturing.

\begin{figure}[!htp]
\centering
\includegraphics[width=0.45\textwidth]{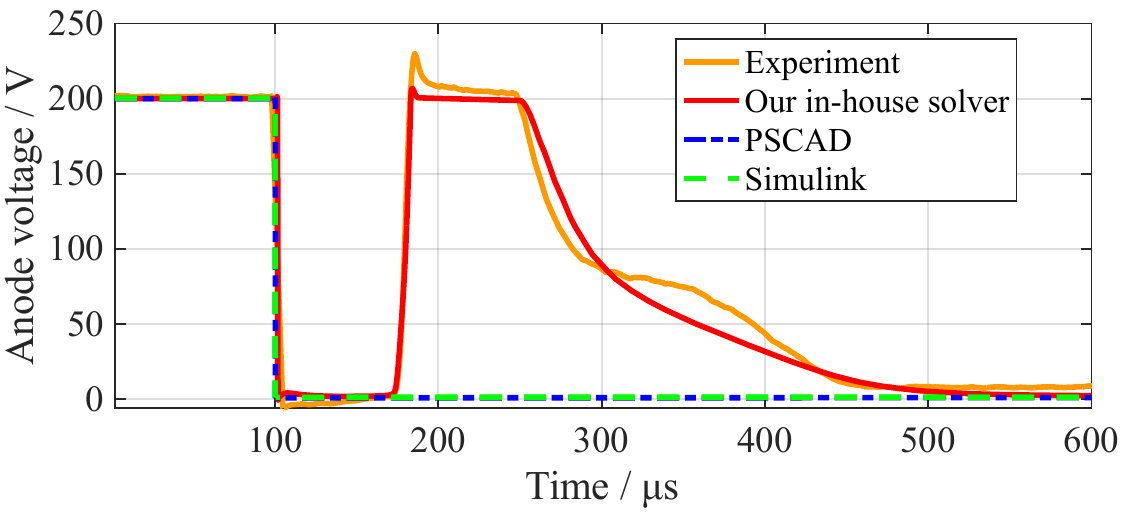}
 \vspace{-10pt}
\caption{Comparison with experiment and circuit-based solvers.}
\label{fig:turn_on_trouble_compare_simulink}
\end{figure}

As shown in Fig. \ref{fig:turn_on_trouble_compare_simulink}, circuit-based solvers, PSCAD/EMTDC and Simulink, falsely predict a normal turn-ON process, failing to capture the observed failure phenomenon, because circuit-based solvers lack the capability to model internal carrier dynamics and the complex interactions between cells.

\begin{figure}[!htp]
\centering
\includegraphics[width=0.45\textwidth]{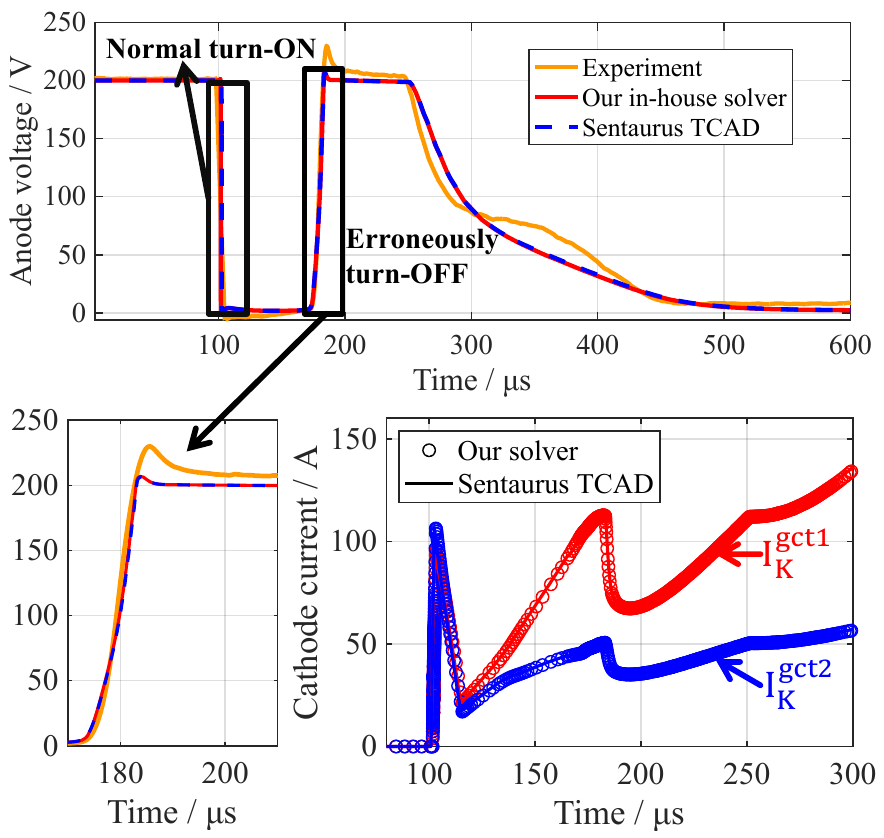}
 \vspace{-10pt}
\caption{Comparison with experiment and multi-physics solver.}
\label{fig:turn_on_trouble_compare_TCAD}
\end{figure}

Alternatively, the results of the HPC framework align closely with those obtained from the commercial TCAD solver within numerical error, as shown in Fig. \ref{fig:turn_on_trouble_compare_TCAD}, and agree satisfactorily with the experimental waveform. The discrepancies observed between the experimental measurements and simulations could be attributed to the simplification in the dual-cell modeling, electromagnetic interference, device deviations, and other factors.

The key stages of this specific failure, as analyzed using the HPC framework and PDAE model, are outlined as follows (cf. Fig. \ref{fig:abnormal_turn_on_detail_analysis} and \ref{fig:loop_current}):

\begin{figure}[ht!]
\vspace{-8pt}
\centerline{
\includegraphics[width=0.45\textwidth]{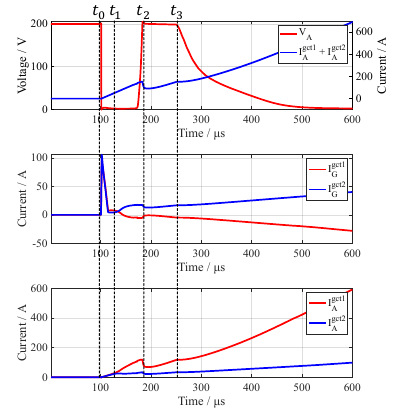}
}
\vspace{-16pt}
\caption{Comparison of the Anode voltage and Anode, Gate currents of the two cells.}
\label{fig:abnormal_turn_on_detail_analysis}
\end{figure}

\begin{figure}[ht!]
\vspace{-12pt}
\centerline{
\includegraphics[width=0.46\textwidth]{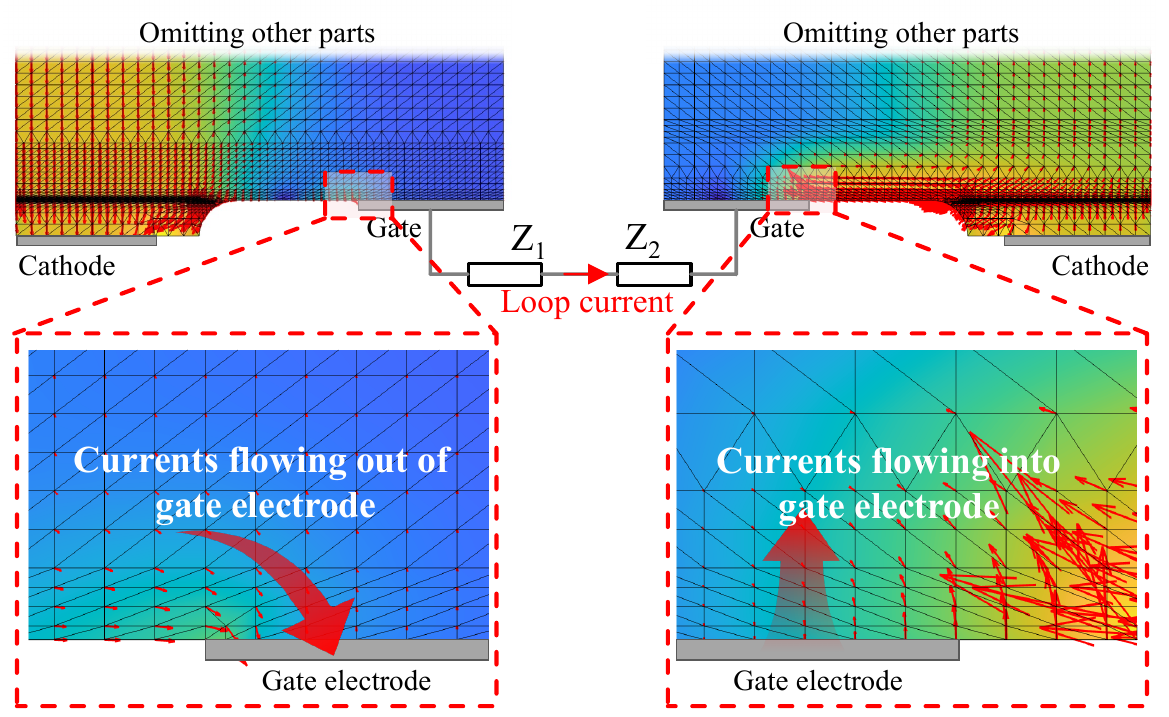}
}
\vspace{-6pt}
\caption{Loop current through the dual-cell experiment at $t_2$ of Fig. \ref{fig:abnormal_turn_on_detail_analysis}, where current densities at each mesh vertices are shown by red arrows.}
\label{fig:loop_current}
\end{figure}

\begin{enumerate}
    \item From $t_0$ to $t_1$: Both cells turn on normally with gate trigger current.

    \item From $t_1$ to $t_2$: Due to the difference in the p$^+$-base region between the cells, a voltage difference occurs at their Cathode-Gate junction (J3 junction). This difference induces a loop current shown in Fig. \ref{fig:abnormal_turn_on_detail_analysis} and Fig. \ref{fig:loop_current}, where the size and direction of red arrows indicate the amplitude and direction of current densities. This difference has two negative effects: (1) reducing the back porch current in GCT1 and (2) decreasing the Anode current in GCT2. These changes cause both cells to gradually lose their latch-up mechanism, leading to an erroneous turn-off at $t_2$.
    
    \item From $t_2$ to $t_3$: As the loop current grows, GCT2 is retriggered. The higher Anode-Cathode voltage and increased Anode current also re-trigger GCT1, causing both cells to turn on again at $t_3$ and remain in the on-state as the Anode current continues to increase.
\end{enumerate}

The above mechanism interprets the divergence in the cathode current observed in Fig. \ref{fig:turn_on_trouble_compare_TCAD} that is connected to an unexpected turn-OFF event. Such type of failures, driven by carrier-level behaviors, is difficult to be analyzed using methods other than PDE-modeled device simulations, which highlights the necessity of PDAE-modeled simulations.

\section{Conclusion}
In this paper, we propose a hybrid-parallel collaborative simulation framework designed for the analysis of power electronic equipment. This framework integrates both device physics and circuit dynamics, thereby effectively bridging the gap between accuracy and efficiency.

The power electronic equipment is modeled using partial differential algebraic equations (PDAEs) that are then decoupled into PDEs and DAEs through dynamic iterations. This crucial technique effectively integrates the device physics with the circuit dynamics. From a circuit perspective, the coupling between the semiconductor device and the remaining part of the circuit is modeled by equivalent conductance and current sources. 

The modeling is followed by hybrid parallel acceleration. This parallel paradigm allocates partitioned computing tasks to different processes and threads on both shared and distributed memory. It leverages the physical connectivity of power electronic equipment and significantly increases the computing speed on clusters with high parallel efficiency.

Demonstrated by three practical experiments, the High-Performance Computing (HPC) framework demonstrates significant potential in the analysis and optimization of power electronic equipment that demands the integration of device physics and circuit dynamics. It stands out in terms of both efficiency and accuracy when it comes to real-world equipment optimization and failure analysis, outperforming conventional commercial simulators which may struggle to yield timely results. 

Additionally, we remark that detailed microscopic parameters, such as doping concentrations, are critical for PDE-based simulations but are not always fully accessible. Although incomplete data may compromise absolute accuracy, these simulations still provide valuable trends and insights for guiding the design and optimization of devices and equipment. Our framework is particularly suitable for collaborative engineering scenarios in which device manufacturers and equipment designers work closely together.

It is possible for this framework to be integrated into existing
SPICE platforms to enhance their multi-physics capabilities,
enabling detailed carrier-level device dynamics to complement
traditional circuit-level simulations. However, challenges such as
parallelization synchronizations must be addressed properly.
We leave this topic for future work.



\vfill

\end{document}